\documentclass[a4paper]{jpconf} 
\usepackage{graphicx}
\usepackage{hyperref}
\usepackage{natbib} 
\usepackage{amsmath}
\usepackage{amssymb}
\begin{document}
\title{Unveiling Axion Signals in Galactic Supernovae with Future MeV Telescopes}

\author{Zhen Xie$^{1,2,3}$, Jiahao Liu$^{1,2,3}$, Bing Liu$^{4,3}$, Ruizhi Yang$^{1,2,3,*}$}
\address{1.Deep Space Exploration Laboratory, Hefei,Anhui 230088, China}
\address{2.CAS Key Laboratory for Research in Galaxies and Cosmology, Department of Astronomy, School of Physical Sciences,\\
University of Science and Technology of China, Hefei, Anhui 230026, China}
\address{3.School of Astronomy and Space Science, University of Science and Technology of China,\\ Hefei, Anhui 230026, China}
\address{4.Key Laboratory of Dark Matter and Space Astronomy, Purple Mountain Observatory, Chinese Academy of Sciences, Nanjing 210023, China}
\ead{*yangrz@ustc.edu.cn}

\begin{abstract}
Axion-like particles (ALPs) produced via the Primakoff process in the cores of Galactic core-collapse supernovae (SNe) could convert into MeV-energy $\gamma$-rays through interactions with the Milky Way’s magnetic field. To evaluate the detection prospects for such signals, we perform sensitivity projections for next-generation MeV telescopes by combining hypothetical instrument responses with realistic background estimates. Our analysis incorporates detailed simulations of the expected ALP flux from nearby SNe, the energy-dependent conversion probability in Galactic magnetic fields, and the telescope’s angular/energy resolution based on advanced detector designs. Background components are modeled using data from current MeV missions and extrapolated to future sensitivity regimes.

Our simulations demonstrate that next-generation telescopes with improved effective areas and energy resolution could achieve sensitivity to photon-ALP couplings as low as $g_{a\gamma} \approx 1.61 \times 10^{-13}~\mathrm{GeV}^{-1}$
for ALP masses $m_a \lesssim 10^{-9}~\mathrm{eV}$ in the Galactic Center. These results indicate that future MeV missions will probe unexplored regions of ALP parameter space, with conservative estimates suggesting they could constrain $g_{a\gamma}$ values two orders of magnitude below current astrophysical limits. Such observations would provide the most stringent tests to date for axion-like particles as a dark matter candidate in the ultra-light mass regime.
\end{abstract}
\noindent{\it Keywords\/}:Dark Matter, Axion-Like Particles, MeV Telescopes, Supernova
\newpage
\section{Introduction}
The axions, as a type of weakly interacting light mass particles, are characterized by a light mass and extremely weak coupling to the Standard Model, emerging as an elegant solution to the strong-CP problem \cite{abbott1983cosmological,dine1983not,Preskill1983, Peccei1977}. Axion-like particles (ALPs), which share similar properties with axions, naturally arise in string theory and are often considered in cosmological models addressing dark energy \cite{raffelt2011mev}. Both axions and ALPs have gained significant attention for their potential role in explaining unsolved mysteries in physics, such as the nature of dark matter and dark energy\cite{Preskill1983, Sikivie1983, Raffelt1996,meyer2017fermi,Calore_2024,Hoof_2023}.

The Primakoff process is a mechanism by which axions or ALPs, interact with photons in the presence of an external electric or magnetic field. In this process, a photon can convert into an axion (or vice versa) as it interacts with a strong electromagnetic field, such as those found in stellar environments or galactic magnetic fields \cite{PhysRev.81.899}. This interaction is central to many astrophysical and laboratory searches for axions and ALPs, as it provides a potential pathway for detecting these elusive particles through their indirect effects on photon behavior \cite{PhysRevLett.51.1415}. The Primakoff process thus serves as a foundational principle in the search for weakly interacting particles like axions, linking their theoretical properties to observable signals in high-energy astrophysics and cosmology \cite{Peccei1977}.

The next generation of space-based telescopes is poised to significantly enhance our ability to detect ALPs by observing photon-ALP conversions, such as those occurring through the Primakoff effect in high-energy astrophysical environments \cite{meyer2017fermi}.
MeV telescopes, such as the upcoming generation of detectors including e-ASTROGAM \citep{deangelis2018science}, AMEGO-X \citep{amego-x}, COSI \citep{tomsick2019compton}, and MeGaT \footnote{Zhang et al. Megat: A High-Resolution MeV Gamma Telescope using TPC and CZT Detectors. Private Communication, 2023.},
will be uniquely suited for this task. These observatories will be capable of capturing high-energy photon signals from astrophysical sources like supernovae (SNe), gamma-ray bursts, and active galactic nuclei, where strong magnetic fields can facilitate photon-to-ALP conversions via the Primakoff process. As these particles travel through space, they may interact with cosmic magnetic fields, leading to the reconversion of ALPs back into photons. This reconversion could produce a distinct signature, allowing for the detection of ALPs by these advanced telescopes \cite{meyer2017fermi, Calore_2024}. The ability to detect such conversions would provide a promising window into the existence of ALPs, which are otherwise too weakly interacting to be directly observed in laboratory experiments.

This paper presents a comprehensive analysis of the potential for detecting axion-like particles using next-generation space-based telescopes sensitive to the MeV energy range. Our approach relies on modeling photon-ALP conversion processes in the strong magnetic fields of various astrophysical environments, such as SNe. Chapter 2 introduces the theoretical framework for ALPs, reviewing their properties and the Primakoff process that enables photon-ALP conversion in SNe. Chapter 3 focuses on the observational capabilities of MeV telescopes, specifically highlighting how these advanced instruments can detect photon-ALP conversions and the expected signatures. Finally, the last Chapter discusses the implications of our findings, providing an outlook on how future observations could help constrain the axion parameter space and potentially reveal the presence of these elusive particles. The results presented here aim to set the stage for upcoming experiments and pave the way for a deeper understanding of dark matter and fundamental physics.

\section{ALP Production Mechanisms in SN}
In an SN core, ALPs can be generated primarily through the Primakoff process \cite{PhysRev.81.899}. In this process, thermal photons are converted into ALPs as they interact with the electrostatic fields of charged particles, such as ions, electrons, and protons present in the dense stellar plasma \cite{Raffelt1996}. To calculate the rate of ALP production via the Primakoff process within an SN core, we closely follow the approach outlined in Ref.~\cite{Payez_2015}.
The differential ALP production rate per unit volume and photon energy is given by:
\begin{align}
    \frac{d\dot{n}_a}{dE} = \frac{g_{a\gamma}^2 g^2_{eff} T^3 E^2}{8 \pi^3 (e^{E/T} - 1)} \left[ \left( 1 + \frac{\xi^2 T^2}{E^2} \right) \ln\left(1 + \frac{E^2}{\xi^2 T^2}\right) - 1 \right],
\end{align}
where \( E \) represents the photon energy, \( T \) is the temperature of the SN core, and \( g_{eff} \) denotes the effective coupling constant that depends on the plasma environment. The parameter \( g_{a\gamma} \) denotes the axion-photon coupling constant, which characterizes the interaction strength between the axion field \( a \) and the electromagnetic field via the term \( g_{a\gamma} a\, \mathbf{E} \cdot \mathbf{B} \) with dimensions of inverse energy.
Here, \( \xi^2 = \kappa^2 / 4T^2 \) with \( \kappa \) being the inverse Debye screening length, accounting for the finite range of the electrostatic field surrounding charged particles within the plasma.

To determine the total ALP production rate, this expression can be integrated over the photon energy across the relevant energy spectrum within the SN environment. This integration yields the overall ALP production per unit volume, providing an estimate of the ALP flux generated in a typical SN explosion. The calculated ALP flux is essential for understanding the potential observational signatures of ALPs in high-energy astrophysical contexts and has implications for dark matter research \cite{Raffelt1996}.

\begin{figure}
    \centering
    \includegraphics[width=0.6\linewidth]{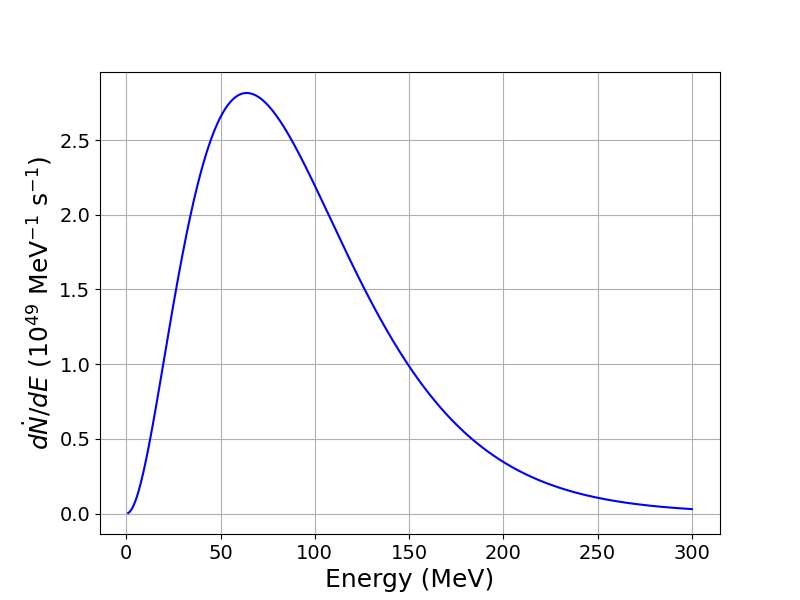}
    \caption{The ALP production rate per unit energy in case of a nearly massless ALP with \( g_{a\gamma} = 10^{-10} \ \text{GeV}^{-1} \), integrated over the explosion time of \( 10 \ \text{s} \).}
    \label{fig:dnde}
\end{figure}

An optimal fit for the total production rate is provided by the expression widely used in SN neutrino studies\cite{Payez_2015,andriamonje2007improved}:
\begin{align}
    \frac{dN_a}{dE} = C \left( \frac{g_{a\gamma}}{10^{-11}\, \mathrm{GeV}^{-1}} \right)^2
    \left( \frac{E}{E_0} \right)^\beta
    \exp\left(-(\beta + 1) \frac{E}{E_0} \right),
\end{align}
shown in Fig.\ref{fig:dnde}. In this expression, \( C \) is a normalization constant, and the parameter \( E_0 \) corresponds to the average energy, \(\langle E_a \rangle = E_0\). We use that $ C = 5.32 \times 10^{50}  ~\text{MeV}^{-1} $, \( E_0 = 94 \ \text{MeV}\), and \( \beta = 2.12 \) for the curve integrated over the explosion time of \( 10 \ \text{s} \) when the progenitor mass about 10 $M_\odot$ as described in Ref.\cite{meyer2017fermi}.

From this, the gamma-ray flux observed on Earth can be calculated as:
\begin{align}
    \frac{dN_{\gamma}}{dE} = \frac{1}{4 \pi d^2} \frac{dN_a}{dE} P_{a\gamma},
\end{align}
where d is the distance of SN, and \(P_{a\gamma}\) is the ALP-photon conversion probability. Given the ALP mass $m_a$, combined with the conversion probability obtained through the magnetic field model, we can constrain $g_{\alpha\gamma}$ using the flux of SN explosions observed on Earth\citep{Horns_2012}.

For ultralight ALPs with masses \( m_a \lesssim 10^{-9} \ \text{eV} \), the strongest existing constraint on \( g_{a\gamma} \) is derived from the absence of observed gamma rays from SN 1987A\cite{Payez_2015}, a core-collapse SN that exploded in the Large Magellanic Cloud at a distance of approximately 50 kpc. 

Building upon the constraints from SN 1987A, recent research\cite{Carenza_2025} has expanded our understanding of axion properties, particularly focusing on both ultralight and massive axions. Ultralight axions, with masses around $10^{-6}$ eV, are of significant interest due to their potential role as dark matter candidates. Their interactions with photons can lead to observable effects in stellar environments, such as enhanced cooling rates in white dwarfs and red giants. These phenomena provide stringent constraints on the axion-photon coupling strength\cite{Payez_2015,meyer2017fermi,Hoof_2023}. And the gamma-ray spectrum from ALP conversions offers a distinctive probe into the pion content of a protoneutron star (PNS)\cite{Lella_2024}. A spectral peak around 200 MeV, resulting from axion-pion interactions, would indicate a significant pion abundance in the PNS core—information not accessible through neutrino observations alone.

On the other hand, massive axions, with masses in the keV to MeV range, can be produced in high-temperature astrophysical settings like SN cores. For axion masses $m_a\geq50~\rm MeV$, the dominant production mechanism in supernovae shifts from nucleon-nucleon bremsstrahlung to photon coalescence, also known as the inverse decay process\cite{PhysRevD.62.125011}. In this process, two photons within the dense and hot SN core can annihilate to produce an axion\cite{raffelt2007axions,PhysRevD.101.103016}. This mechanism becomes particularly significant for heavier axions, as the production rate via photon coalescence can surpass that of the Primakoff process by orders of magnitude in this mass range, and applied to muonphilic bosons\cite{PhysRevD.105.035022} and generic $e^+e^-$ decays\cite{PhysRevD.105.063026}. If these axions decay into photons, they could generate detectable gamma-ray signals. Detection or non-dectection of such such signals from events like SN 1987A provides constraints on the properties of massive axions
\cite{Rembiasz_2018,Mori_2022,Berezhiani_2000,PhysRevLett.128.211101,mori2023multimessengersignalsheavyaxionlike}.

\section{Background and Sensitivity Calculations}
Recently, several next-generation MeV detector projects, such as e-ASTROGAM \citep{deangelis2018science}, AMEGO \citep{mcenery2019amego}, COSI \citep{tomsick2019compton}, and MeGaT, have demonstrated significant improvements in sensitivity compared to current MeV instruments.
Instead of using specific instrument response data, we assume that future MeV instruments will have an effective area of $100~\rm cm^2$ and a point spread function of $2^\circ$. These assumptions are consistent with the design and preliminary simulation outcomes for both semiconductor detectors, such as e-ASTROGAM, and gas detectors, such as MeGaT.

We selected four candidates as potential study objects for SN explosions as selected by Fermi-LAT work\cite{meyer2017fermi}. The first is Betelgeuse, also known as Alpha Orionis, which is a prominent red supergiant star located in the constellation Orion. It is one of the brightest stars visible in the night sky and is easily recognizable as the top left "shoulder" of the Orion constellation. Betelgeuse is notable for its immense size, with a diameter estimated to be over 700 times that of the Sun, making it a stellar giant in the late stages of its life cycle. Due to its advanced evolutionary state, Betelgeuse is expected to end its life in a dramatic SN explosion, an event that would be visible from Earth even during the daytime. The star’s variability in brightness, which has intrigued astronomers for centuries, results from complex processes within its outer layers. Betelgeuse’s study offers valuable insights into the life and death of massive stars\cite{carroll2006introduction,harper2008betelgeuse,levesque2009properties}. Due to its proximity and potential for a SN explosion, we consider this target an excellent sample for estimating the constraints that next-generation MeV detectors can place on axion particles.

We also considered M31, the Andromeda Galaxy, which is the closest spiral galaxy to the Milky Way and one of the most studied galaxies due to its proximity and similarity in structure. Located approximately 778 kpc away, M31 offers a valuable opportunity to study stellar evolution, galactic dynamics, and potential SN events. Historically, SNe in M31 have been rare, but its massive stellar population suggests the potential for future core-collapse or Type Ia SNe. Observations and research on M31 help refine models of stellar death and the distribution of such explosive events in spiral galaxies\cite{binney2008galactic}. To clarify the detector's sensitivity and compare it with related work, we also calculated the cases at the GC and SN 1987A positions.

\begin{table}[h!]
\centering
\begin{tabular}{l|cccc}
 & Betelgeuse & M31 & SN 1987A & GC\\
\hline
R.A. (°) & 88.79 & 10.63 & 83.87 & 266.42\\
Dec. (°) & 7.41 & 41.30 & -69.27 & -28.99\\
Distance (kpc) & 0.197 & 778 & 51.4 &8.5\\
\hline
\end{tabular}
\caption{Basic information for Betelgeuse, M31, SN 1987A and GC.}
\label{tab:parameters}
\end{table}

\begin{figure}[h]
    \centering
    \includegraphics[width=0.6\linewidth]{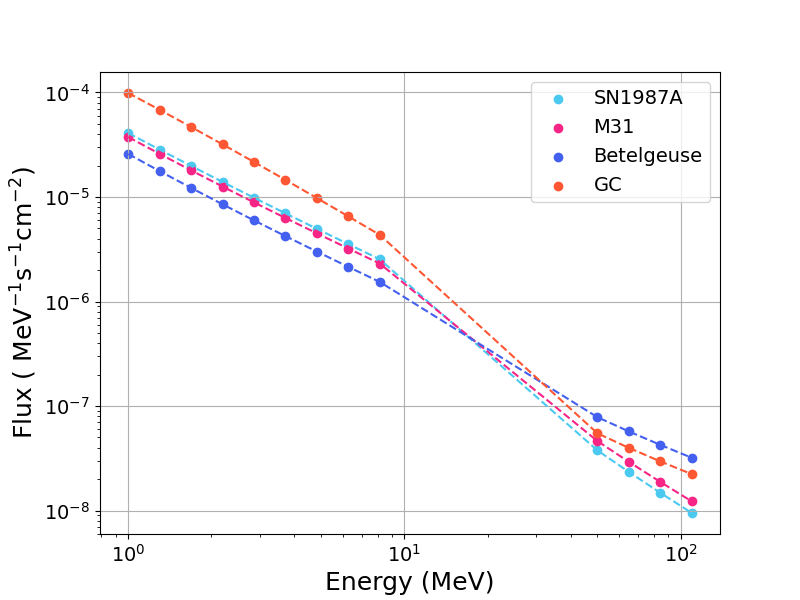}
    \caption{Background flux in SN1987A, Betelgeuse and M31 region with a radius of 2 degrees, the background of 1-10 MeV is extrapolated from Ref.\cite{Siegert_2022}, and 50-100 MeV is from the work of Fermi-LAT\cite{fermibkg}. The dashed line is the function we use to calculate the background by interpolating these data.}
    \label{fig:enter-label}
\end{figure}

To estimate the sensitivity of the instrument to the explosion, we need to calculate the background gamma-ray intensity in the region of interest\cite{PhysRevD.109.043020}. The diffuse emissions below $10~\rm MeV$ near the Galactic center were recently reanalyzed by Ref.\cite{Siegert_2022} using INTEGRAL/SPI observations. To estimate the background at different sky positions, we extrapolated these results, assuming that the spatial distribution in the MeV band follows the energy-dependent template predicted by GALPROP models \citep{galprop}. For emissions above 50 MeV, we used the Fermi-LAT interstellar emission model ({\it gll{\_}iem\_v07.fits})\citep{fermibkg} based on the first 9 years of data, integrating the flux within a specific region of interest. The two background values were smoothly connected by interpolation between $10~\rm MeV$ and $50~\rm MeV$. The results for these regions are illustrated in Fig.\ref{fig:enter-label}.

We selected the energy range of 1--100 MeV for integration to cover the most prominent part of the ALP spectrum as shown in Fig.\ref{fig:dnde}, and chose an exposure time of \( 10 \ \text{s} \) as the most intense phase of an SN explosion\citep{Payez_2015}. Firstly, we estimated the background counts. This is achieved by integrating the background flux over the exposure time and the effective area of the observation instrument. Through calculations, we find that although the background in the Galactic Center is slightly higher than in other regions, the background counts are all less than 1. Using the method outlined in \citep{feldman1998unified}, we can estimate the 95\% confidence interval of the SN signal by setting the background counts to 1, equal to the smallest integer greater than the background signal, as a conservative estimate. Using this method, we obtain a 95\% confidence upper limit of 4.14 counts, representing our MeV detector's capability in detecting SN explosion signals. It can be observed that this result is in close agreement with the Fermi-Lat results\cite{meyer2017fermi}. 

To constrain \( g_{a\gamma} \) in the parameter space, we then follow the method mentioned in \cite{Payez_2015}.
We assume the JF12 galactic magnetic field model\cite{2012ApJ...757...14J} to calculate the $P_{\alpha\gamma}$ in the Interstellar Magnetic Medium, which is a relatively conservative magnetic field model for ALP conversion.
As pointed out in the Fermi-LAT study\cite{meyer2017fermi}, the impact of these parameters on sensitivity estimation is not significant. Therefore, we choose a progenitor mass of 10 $M_\odot$ to calculate the sensitivity as a widely accepted model\cite{neilson2011asp}. We numerically solve for $g_{a\gamma}$  by varying the value of $m_a$  in the parameter space and determine the position of the turning point in the parameter space by identifying the mass range corresponding to a stable $g_{a\gamma}$ value following the technique described in\citep{Horns_2012}. 
It is observed that the constraints of the ALP parameter gradually weaken as the mass increases. 
Through calculations, we find that for masses \( m_a \lesssim 10^{-9} \ \text{eV} \), the sensitivity of MeV detector to \( g_{a\gamma} \) in the GC region can reach \( 1.61 \times 10^{-13} \ \text{GeV}^{-1} \), 
and the sensitivity in the Betelgeuse region is approximately \( 1.26 \times 10^{-13} \ \text{GeV}^{-1} \). 

To account for error bars, we considered the following components. Since we did not employ the on-off method, we only accounted for Poisson fluctuations in the background. For the impact of different Galactic magnetic field models, we referenced results from previous studies on core-collapse SNe in the Galactic \cite{Calore_2024} and adopted an average error range based on their findings.

In addition to the previously proposed model extending up to 100 MeV, we also consider an alternative scenario in which the detector's effective area is assumed to be non-vanishing only up to 30 MeV, as $m_a$ below 30 MeV correspond to the main production channel for the Primakoff process\cite{Carenza_2025}. This lower energy limit is more representative of the capabilities of current and near-future detectors, which are generally expected to operate effectively within this energy range. In this model, the instrument's performance is assumed to significantly decrease beyond 30 MeV, reflecting the expected limitations in sensitivity at higher energies. By incorporating this more conservative energy threshold, we can better assess the potential performance of detectors optimized for lower-energy observations, providing a broader comparison for experiments targeting energies up to 30 MeV. Due to the fact that the effective area and energy range is not as large as that of Fermi-LAT, the expected results in the Galactic Center in this case will not be as strong as Fermi-LAT's result.

\begin{figure}[h]
    \centering
    \includegraphics[width=0.8\linewidth]{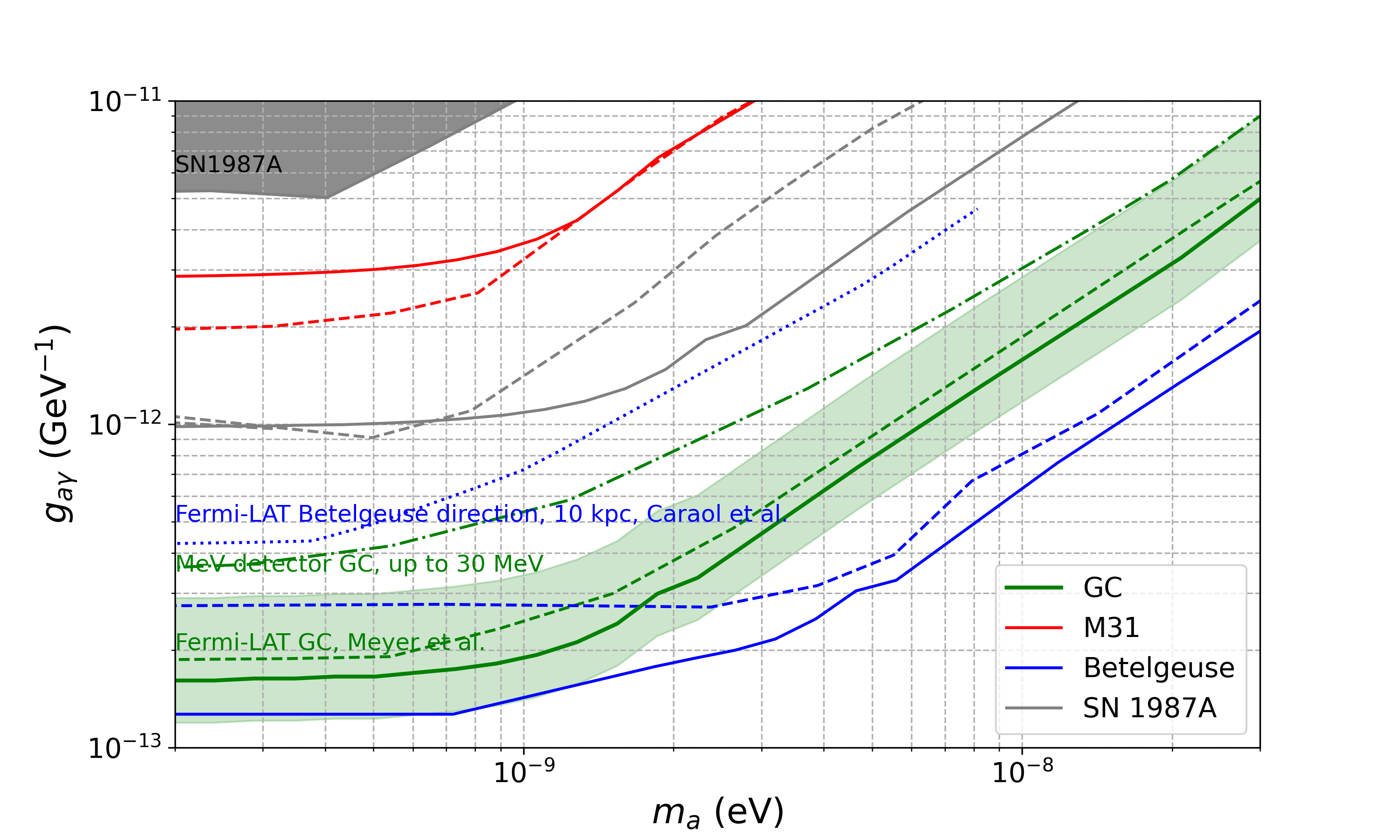}
    \caption{Our best results of expected limits on ALP parameters from an SN explosion. The gray area represents the results from SN 1987A\cite{Payez_2015}, the Fermi-LAT results\cite{meyer2017fermi,Calore_2024} are shown with dashed lines, and solid lines of the same color indicate the MeV detector results. The shaded area shows the error range of the results at the Galactic Center(with solid line), and the dot-dashed line represents the case for an instrument with a maximum energy limit of 30 MeV.}
    \label{fig:result}
\end{figure}

Fig.~\ref{fig:result} presents our results, although the effective area is small, we still derive a stronger sensitivity, since it covers the vast majority of the ALP spectrum. And we present the result where the instrument's energy range extends only up to 30 MeV of the Galactic Center for comparison. Since the MeV detector is still in the conceptual phase, our estimates of background intensity for sensitivity predictions are based on simplified assumptions rather than rigorous data simulations or the “On-Off” method used by Fermi-LAT. This may partially explain the weaker constraints that we obtained for M31 due to its greater distance. However, we believe that with future observational data and the application of more advanced techniques, the sensitivity will improve further, and the current estimates should be regarded as conservative. Even so, our results sufficiently demonstrate the potential of next-generation MeV detectors in ALP research.

It is important to note that this result is only obtained under optimal observational conditions. This means that at the time of an SN explosion, our detector must be perfectly pointed at the relevant region to achieve the maximum sensitivity. One feasible approach is to use neutrino observations to predict the supernova's occurrence, increasing the likelihood of detecting an SN explosion and obtaining timing information of $\gamma$-photons. 

\section{Discussion and Conclusion}
In this study, we analyzed the potential of next-generation MeV detectors to constrain ALP parameters, focusing on regions with potential SN activity. We detailed the selection of energy ranges and outlined the methodology for estimating background gamma-ray flux and calculating sensitivity through numerical solutions. Our approach included integrating the effective area and exposure time to derive the background count rate and establishing detection thresholds for \( g_{a\gamma} \) by analyzing the ALP-photon conversion probabilities. 
It can be seen that, once a SN explosion within the Milky Way is observed, future MeV detectors will be able to place stronger constraints on ALP dark matter. Since the MeV detector can cover the vast majority of the ALP spectrum, such observations offer stronger constraints and can improve our existing particle theories by analyzing the finer details of the spectrum.

Since SN explosion events are rare in the universe, as \citep{diehl2006radioactive,keane2008birthrates,adams2013observing} points out, only 2-3 times per century in the Galaxy. Increasing the effective area of the MeV telescope may not significantly increase the probability of observing this event while more background counts will be introduced. As mentioned at the end of the last chapter, a possible approach is using neutrino observations as a harbinger of an SN explosion. Neutrinos can provide an early warning of impending events as products of the early stages of an SN explosion. This early signal would give us a crucial early warning. 
During a SN explosion, neutrinos are emitted in a brief and intense burst over approximately 10 seconds as the stellar core collapses. These neutrinos escape the dense core almost instantly due to their weak interaction with matter, reaching Earth before the associated light signals. The light from the explosion emerges when the sub-lightspeed shockwave, traveling at approximately 50\% of the speed of light, energizes and illuminates the outer stellar envelope. This delay, ranging from minutes to hours, provides a temporal window in which neutrinos act as the first detectable messengers.

Existing studies indicate that a Galactic SN can indeed be located by its neutrino signal\cite{PhysRevD.60.033007,Tom_s_2003,mukhopadhyay2020presupernova,syvolap2024highenergyneutrinosignalssupernova}, with the most effective method being neutrino-electron scattering in a water Cherenkov detector. Detectors such as SK and Sudbury Neutrino Observatory (SNO) are capable of performing this measurement independently. Based on these results, under normal circumstances, neutrino detectors can constrain the sky area of a SN burst to within several degrees. Considering the low spatial density of potential SN candidates in the sky\cite{green2024updatedcatalogue310galactic} and accounting for the instrument's field of view we considered($2^\circ$), this sparse distribution allows us to precisely target known candidate regions or directly observe these areas while effectively minimizing interference from other candidates. We conclude that the possibility of observing a SN explosion remains high.

A recent joint study by the KamLAND and Super-Kamiokande(SK) experiments\cite{abe2024combined} resulted in a pre-SN alert system that can detect neutrino signals from a 15 $M_\odot$ star within 510 pc, with a false alarm rate of less than once per century. For stars like Betelgeuse, this system can provide early warnings up to 12 hours before a SN event. These findings highlight the critical role of neutrino detectors in providing early alerts for SN explosions, offering valuable insights into the dynamics of such celestial phenomena.

Several neutrino detectors are already operational or planned for deployment soon. Existing facilities such as SK\cite{santos2024superk} and IceCube\cite{kurahashi2023icecube} have demonstrated their ability to detect neutrinos from various astrophysical sources, including SNe. Upcoming detectors like Hyper-Kamiokande\cite{di2017hyper}, JUNO (Jiangmen Underground Neutrino Observatory)\cite{juno2024potential}, and DUNE (Deep Underground Neutrino Experiment)\cite{abi2020volume} promise even greater sensitivity and broader coverage. And about $10^4$ neutrino events can be detected by Super-Kamiokande from an SN explosion event in 10 kpc\cite{scholberg2012supernova}. These advanced detectors will enhance our ability to capture neutrinos from the pre-SN phase and can also be used as the time information of the event to provide a benchmark for data analysis, contributing to a deeper understanding of the mechanisms leading to SN explosions.

To conclude, our results provide useful insights into the design and potential of future MeV missions for investigating ALP properties. If we are "fortunate" enough to observe an SN explosion in the future, MeV detectors will undoubtedly provide strong constraints on ALP dark matter.

\section{Acknowledgment} 
Ruizhi Yang is supported by the NSFC under grant 12393854 and 12041305, and gratefully acknowledges the support of Cyrus Chun Ying Tang Foundations and the studio of Academician Zhao Zhengguo, Deep Space Exploration Laboratory.
Bing Liu acknowledges the support from the NSFC under grant 12103049.
\appendix

\bibliography{main}

\end{document}